\documentclass[manuscript,screen]{acmart}

\usepackage{multirow}

\usepackage[flushleft]{threeparttable} 
\usepackage{booktabs,caption}

\newboolean{showcomments}
\setboolean{showcomments}{true} 
\ifthenelse{\boolean{showcomments}}
  {\newcommand{\nb}[2]{
    \fcolorbox{gray}{yellow}{\bfseries\sffamily\scriptsize#1}
    {\sf\small\textit{#2}}
   }
   
  }
  {\newcommand{\nb}[2]{}
   
  }

\newcounter{quotecount}

\usepackage[normalem]{ulem} 
\usepackage{xcolor}

\newcommand{\ins}[1]{#1} 
\newcommand{\del}[1]{} 
\newcommand{\chg}[2]{#2} 

\newcommand{\delmm}[1]{} 

\acmConference[SE 2030]{International Workshop on Software Engineering in 2030}{November 2024}{Puerto Galinàs (Brazil)}

\AtBeginDocument{%
  \providecommand\BibTeX{{%
    \normalfont B\kern-0.5em{\scshape i\kern-0.25em b}\kern-0.8em\TeX}}}

\setcopyright{acmlicensed}
\copyrightyear{2018}
\acmYear{2018}
\acmDOI{XXXXXXX.XXXXXXX}

\acmJournal{JACM}
\acmVolume{37}
\acmNumber{4}
\acmArticle{111}
\acmMonth{8}

\acmISBN{978-1-4503-XXXX-X/18/06}

\begin{document}

\setcounter{section}{0}
\setcounter{page}{1}


\title{Automation in Model-Driven Engineering: A look back, and ahead}

\author{Lola Burgue\~no}
\authornote{All the authors contributed equally to this research.}
\orcid{0000-0002-7779-8810}
\affiliation{%
  \institution{ITIS Software, University of Malaga}
  \country{Spain}
}
\email{lolaburgueno@uma.es}

\author{Davide Di Ruscio}
\authornotemark[1]
\affiliation{%
  \institution{University of L'Aquila}
  \country{Italy}
}
\email{davide.diruscio@univaq.it}

\author{Houari Sahraoui}
\authornotemark[1]
\orcid{0000-0001-6304-9926}
\affiliation{%
  \institution{DIRO, Université de Montréal}
  \country{Canada}
}
\email{sahraouh@iro.umontreal.ca}

\author{Manuel Wimmer}
\authornotemark[1]
\orcid{0000-0002-1124-7098}
\affiliation{%
  \institution{Johannes Kepler University Linz}
  \country{Austria}
}
\email{manuel.wimmer@jku.at}

\renewcommand{\shortauthors}{Burgue\~no, Di Ruscio, Sahraoui, and Wimmer}

\begin{abstract}

Model-Driven Engineering (MDE) provides a huge body of knowledge of automation for many different engineering tasks, especially those involving transitioning from design to implementation. With the huge progress made in Artificial Intelligence (AI), questions arise about the future of MDE, such as how existing MDE techniques and technologies can be improved or how other activities that currently lack dedicated support can also be automated. However, at the same time, it has to be revisited where and how models should be used to keep the engineers in the loop for creating, operating, and maintaining complex systems. To trigger dedicated research on these open points, we discuss the history of automation in MDE and present perspectives on how automation in MDE can be further improved and which obstacles have to be overcome in both the medium and long-term. 

\end{abstract}

\maketitle

\section{Introduction}

B\'ezivin defined the term Model-Driven Engineering (MDE) in 2005 as follows: ``\textit{MDE is a software engineering methodology which uses formal models, i.e., models which are machine-readable and processable, to produce executable software systems semi-automatically}''~\cite{Bezivin05}. The main idea behind \textit{formal models} is to provide abstractions of software systems to manage their complexity. Based on these abstractions, as they are machine-readable and even processable, \textit{automation techniques} are provided, e.g., to generate executable software~\cite{Schmidt06}, which was the main focus 20 years ago in the pioneering phases of MDE. However, other automation concerns soon emerged and have been studied, such as analysing models by transforming them into formal domains such as model checkers. Model-to-model transformations have been used for this purpose. Model-to-code transformations have been developed to automate the transition from design models to code-based implementations. To complete the picture, code-to-model transformations have been used to reverse engineer existing software systems such as legacy systems. A comprehensive study on different transformation intents is provided in~\cite{LucioADLSSSW16}. 

Soon the community realized that transformations are complex software systems themselves~\cite{SendallK03}. Hence, significant research efforts have been spent developing model transformation languages to develop automation support in terms of transformations. In parallel, language engineering approaches based on metamodeling techniques~\cite{AtkinsonK03a} have been emerging, which allow for the efficient and effective development of modeling languages and accompanying modeling tool support such as model editors, validators, and storage facilities. These infrastructures enabled the development of domain-specific modeling languages (DSMLs)~\cite{ChenSN05} in addition to general-purpose modeling languages such as UML.  

By having abstractions as models combined with automation support, the promise of MDE is to allow engineers to focus on the problem domain, i.e., which systems should be realized, rather than on the solution domain, i.e., how the system is realized.
\chg{However, several challenges have to be tackled to achieve this promise, as reported over the years in numerous papers.}{Achieving this promise involves addressing several challenges, which have been discussed in various studies.}
In~\cite{FranceR07}, the authors discuss challenges associated with modeling languages, e.g., how to engineer DSMLs, achieving separation of concerns to deal with complexity, and model manipulation/management mostly supported by transformations. In~\cite{MussbacherABBCCCFHHKSSSW14}, the authors discuss how MDE may be used even beyond software, raising the need for interdisciplinary modeling support, e.g., for socio-technical systems. The latter requires model integration, consistency, and strong tool support. Finally, in a more recent work~\cite{BucchiaroneCPP20}, the authors discuss important aspects of MDE, such as foundational, social, community, domain, and tool aspects. However, we are not aware of any report explicitly focusing on the automation aspect of MDE and how this aspect will develop toward 2030. It is important to close this gap as with the recent and expected advances of AI, the current automation techniques and technologies of MDE are highly impacted as well, and new ones may be developed to support additional tasks in the software engineering process that are not targeted by existing technologies. Thus, a discussion on the role of models, MDE methodologies, and engineers in this new era is required to frame the potential future of MDE. This is also of paramount importance for the further adoption of MDE in different domains. Previous studies have reported successful adoptions of MDE for particular domains \ins{(e.g., see~\cite{vallecillo2015industrial,LiebelMTLH18,NISTALA2024103058,DANIEL2024103032,BUSCHHAUS2024103033,SADOVYKH2024103047,JAMES2024103057,KURTEV2024103067})}.  However, challenges and needs for the adoption of MDE have been reported~\cite{hutchinson2014model,HoltmannLS24} recently, which may be tackled with improved automation capabilities such as improving the interoperability of MDE tools as well as providing sophisticated model management support to allow for agile processes. Moreover, for successful adoption, it is essential to rely on high-quality models (in terms of completeness, correctness, and consistency with the systems realized) that evolve to reach high automation levels~\cite{HoltmannLS24}. Of course, reaching these high-quality levels requires additional efforts, which may be mitigated by the application of AI in the future. 

To tackle these challenges, important questions have to be answered, such as which phases can be additionally automated, e.g., should we directly go from natural language requirements to code, or is there still, mainly because of the enhanced AI capabilities, a sweet spot for using models in such highly automated processes? Can we now ease and automate the modeling process itself and reach, at the same time, a higher degree of automation in the synthesis process, such as code generation? These questions are important for the foundations of MDE, but at the same time, also for recent application areas such as low-code/no-code development~\cite{RuscioKLPTW22} and digital twin engineering~\cite{tao2018digital}.    

\ins{This article pursues two primary objectives. The first is to conduct a retrospective analysis of how automation needs in Model-Driven Engineering (MDE) have historically been addressed through the application of various enabling technologies. The second objective is to explore future perspectives on MDE automation. These forward-looking considerations are influenced by two key factors. First, the evolution and transformation of existing paradigms, such as the rapid advancements in machine learning and artificial intelligence, are redefining the potential for automation. Second, the emergence of novel paradigms, like quantum computing or neuromorphic computing, presents unprecedented opportunities to rethink the scope and scale of automation in MDE. Additionally, the growing emphasis on non-functional concerns, including security, sustainability, and ethical considerations, introduces new dimensions to how automation solutions are envisioned and implemented.
It is important to highlight that future perspectives on automation are not merely responses to the limitations of previous approaches. Instead, they are driven by the creation of new opportunities arising from the continuous evolution of technology and the emergence of cutting-edge paradigms. This dual focus on past achievements and future possibilities aims to inspire innovative directions for advancing automation in MDE.}

\del{In the following, we outline the impact of automation in MDE by listing key engineering activities that have been the focus of automation, as well as the techniques used to achieve them. Based on this foundation, we lay out several perspectives for automation in MDE as well as challenges which have to be tackled in order to achieve further automation potentials until 2030.}  

\ins{To achieve these objectives, we began by surveying key activities in MDE that require automation (Section~\ref{automation-activities}). Subsequently, we identified the primary technologies that have enabled varying degrees of automation for these activities and gathered illustrative examples of representative work (Section~\ref{enabling-technologies}). While we collected a substantial body of work, it is important to note that we did not conduct a systematic literature review, as this would extend beyond the scope of our article. The third step involved synthesizing our findings by mapping tasks to automation technologies, thereby completing the overall picture (Section~\ref{mapping}).
For the exploration of future automation perspectives until 2030, we examined both challenges and opportunities across four critical dimensions: \textit{advancements in artificial intelligence}, the \textit{role of human factors}, the \textit{diversity of application domains, users, and modeling interfaces}, and the \textit{growing importance of non-functional concerns} such as security and sustainability (Sections \ref{AI-MDE} to \ref{non-functional-concerns}).}
\section{Achievements in Automation}

\ins{In this section, we present the evolution of MDE automation over the years. We divide it into two subsections: Section~\ref{automation-activities} presents the activities and MDE that require automation, and Section~\ref{enabling-technologies} presents the primary technologies that have enabled such automation together with examples of them..}


\subsection{Targeted Automation Activities\protect\footnote{Note that we will refer to activities as \emph{activities} or as \emph{tasks} indifferently.}}
\label{automation-activities}

In this section, we categorize and discuss the activities that have been targeted for automation within the MDE field, organizing them according to the categories outlined in the contents of a Model-Based Software Engineering Body of Knowledge~\cite{BurguenoCFKLMPP19}.
It should be noted that while these basic tasks serve as foundational components, they can be composed into more complex automation pipelines for different purposes.



\subsubsection{Modeling}

The task of modeling itself, i.e., creating models as abstractions of software and systems, has been the target of automation for many years. In particular, specific efforts have been put into approaches for the creation of DSLs and DSVLs, for instance, using frameworks and tools such as \del{Xtext\footnote{\url{https://eclipse.dev/Xtext}},} JetBrains MPS\footnote{\url{https://www.jetbrains.com/mps}} or the Eclipse Modeling Framework (EMF)\footnote{\url{https://eclipse.dev/modeling/emf}}\del{and Sirius\footnote{\url{https://eclipse.dev/sirius}}}.
Modeling tools have also helped improve MDE workflows and reduce the manual effort required. Examples of these tools are Papyrus\footnote{\url{https://eclipse.dev/papyrus}} \del{, Enterprise Architect\footnote{\url{https://sparxsystems.com}},} and  MagicDraw\footnote{\url{https://www.magicdraw.com}} \del{, USE\footnote{\url{https://github.com/useocl/use}}, and Visual Paradigm\footnote{\url{https://www.visual-paradigm.com}}}.

Other modeling activities that over the years have been the target of automation are model generation~\cite{SainiMGK22}, model search~\cite{LopezC20}, model completion~\cite{chaaben2023towards} and model classification~\cite{nguyen2019automated}.
A long-lasting line of work to automate the execution of models, which is worth emphasizing, has been heavily addressed by researchers working on models@runtime~\cite{BencomoGS19}, where the models of a system need to be up-to-date while the system is running at the same time that is used to adapt the runtime behavior. This line of work is a prominent example of having automated modeling in use.

\subsubsection{Model Quality Management}

Significant efforts have been made to automate activities to find defects, faults, and failures in models and modeling processes.
Numerous model validation and verification techniques exist that automate quality checks, such as correct syntax, semantics, structure, behavior, etc., of models to ensure that any inconsistencies or errors are detected and rectified at the model level and as soon as possible. For instance, model checking has been automated through algorithms that check the temporal behavior of a software or system model against certain properties or specifications, ensuring they meet specified requirements or constraints, and there are numerous languages and mature tools available~\cite{spin,uppaal,prism}.

Automated testing techniques based on models (a.k.a. model-based testing) have been developed, too, where models are used to generate test cases automatically (e.g., \cite{Pretschner05}). Traditionally, this method has helped identify potential problems early in the development process, significantly reducing cost and effort in the testing phase. Moreover, in recent years, model-based testing automation has gained relevance due to the rise of digital twins~\cite{BickfordBBP20}, which impose the need for streamlining testing, and the adoption of new development methodologies such as agile, continuous development, integration, and delivery~\cite{JongelingCC19} where the process of executing test cases and analyzing the results needs to be automated in continuous integration/continuous deployment (CI/CD) pipelines. 

\subsubsection{Model-based execution}

It is necessary to have executable models~\cite{kleppe2003mda} and simulation facilities for testing, debugging, or analysis purposes.
In this regard, executable UML models are popular and usually use action languages to define the behavior of the system (e.g., \del{the Alf Action Language~\cite{alf},} SOIL~\cite{SchaferG20}, or  FUML\footnote{\url{https://www.omg.org/spec/FUML}}).

Over the years, the automation of model-based execution, simulation, and co-simulation have seen numerous advancements, moving from manual processes that required human intervention---e.g., to set up simulations and adjust parameters based on expert knowledge and observations---to highly automated processes and tools. For instance, there has been the standardization of an Action Language as part of the OMG UML Specification for the execution of UML models~\cite{alf} and tools for simulation such as Modelica\footnote{\url{https://modelica.org}}, which are widely used by system engineers. There is also available a series of interpreters for modeling languages, both graphical and textual, many of them fully integrated into modeling environments, e.g., as in the case of the USE approach~\cite{GogollaBR07}.

Furthermore, there have been efforts to come up with model execution support by developing dedicated interpreters for domain-specific modeling languages, e.g., see initiatives such as GeMOC (Globalization of Modeling Languages)\footnote{\url{https://gemoc.org}} focusing on the development of technologies and methodologies to generate dedicated execution-oriented tools for models such as model testing and debugging capabilities based on the defined operational semantics of the modeling languages.

\subsubsection{Model Transformations}

Model transformations constitute a fundamental aspect of MDE and serve as the backbone for automating the transition from high-level abstractions to executable code, simulations, or other models. These are perhaps the most exploited automation and automated activity in MDE, where, among other uses, models are automatically transformed into executable code in various programming languages. They can be classified into model-to-model transformations (M2M), model-to-text (M2T) transformations, and text-to-model (T2M) transformations~\cite{kleppe2003mda}. M2M transformations \ins{receive a model as input and provide another model as output (e.g., a BPMN model can be transformed into a Petri Net model)}. M2T transformations provide the ability to autogenerate text from models, contributing significantly to software development automation, where the models serve as the primary artifact and, from them, textual artifacts are automatically generated~\cite{klein_2015} such as code \ins{(e.g., C\# code from UML state machines)}, documentation, or other artifacts (such as JSON files \ins{for data exchange}). T2M transformations are used to reverse engineer code in models, as well as other purposes such as documentation generation~\cite{Wang10} \ins{(for example, to convert legacy code into a software architecture model)}.

At the same time, model transformations have enabled and pushed forward automation in software engineering; they have also been the target of automation. 
This automation leverages specific languages, execution engines, tools, and frameworks, as described below.
One of the primary means of automating model transformations is through dedicated transformation languages. These languages provide constructs that describe how model elements must be mapped to their equivalent elements. Examples of such languages include ATL~\cite{ATL}, QVT~\cite{QVT}, ETL~\cite{ETL}, and Henshin~\cite{HenshingArendtBJKT10} to name just a few---for a survey see~\cite{KahaniBCDV19}.

Transformation engines\footnote{\url{https://projects.eclipse.org/projects/modeling.mmt.qvt-oml}}$^,$\footnote{\url{https://wiki.eclipse.org/ATL/EMFTVM}}\cite{BurguenoWV16,9146715,TisiPC13} are specialized software components that execute the transformation rules defined in a transformation language against source models to generate target ones. These engines manage the complexities of navigating model structures, applying transformation rules, resolving dependencies, and ensuring the integrity of the generated models. Transformation engines can be standalone tools or integrated into broader MDE platforms, providing a seamless experience from model editing to transformation.

Model transformations must be well-integrated with the software development environment for automation to be truly effective. This includes providing user-friendly interfaces for defining and managing transformations.
Such integration enables developers to incorporate model transformations seamlessly into their workflows, automating repetitive tasks and facilitating more agile and iterative development processes. Some efforts include the Epsilon Playground~\cite{KolovosG22} and AToMPM~\cite{SyrianiVMHME13}, among others.





\subsubsection{Model Management}

Model management involves different activities, which we discuss in the following.
%
%
Automated \emph{model evolution} involves strategies for evolving models to reflect changes in requirements or design. Automation can be achieved through predefined transformation rules or templates that apply changes across models, ensuring consistency and reducing manual effort. Languages such as Epsilon~\cite{KolovosG22} or ATL~\cite{ATL} support automated transformations that facilitate model evolution.

When metamodels evolve, models that conform to those metamodels may become invalid and must be adapted to remain consistent with the new version of the metamodel~\cite{HerrmannsdoerferR09}. Automated \emph{model migration} is needed to update models. So far, this has been achieved in numerous ways, for instance, by means of model transformations~\cite{SprinkleK04,DIRUSCIO12} or with dedicated languages such as Epsilon Flock~\cite{RoseKPPP14}.

\emph{Maintaining models} to ensure their accuracy and relevance over time can be automated through continuous integration tools that integrate model changes, run automated validations, and report inconsistencies or errors. Automated testing frameworks for models also contribute to maintenance by ensuring that model updates do not break existing functionalities.

Since models are software artifacts, version control systems are beneficial for keeping up with current development needs. Given the different nature of models w.r.t. code, approaches to deal with models are needed. For instance, EMFStore~\cite{KoegelH10} or Connected Data Objects (CDO)\footnote{\url{https://www.eclipse.org/cdo}} provide automated support for \emph{model versioning}.

Automated \emph{model diffing tools} compare and identify differences between model versions. EMF Compare\footnote{\url{https://eclipse.dev/emf/compare}} is an example of a tool that provides automated support for model comparison, highlighting differences and conflicts within models developed using EMF (Eclipse Modeling Framework)\footnote{\url{https://eclipse.dev/modeling/emf}}.

\emph{Merging model} changes, especially in collaborative environments or when reconciling parallel development efforts, require sophisticated automation to handle conflicts and ensure model integrity. Tools like EMF Compare\footnote{\url{https://eclipse.dev/emf/compare}} also support automated model merging by providing strategies and user interfaces to resolve conflicts and integrate changes from different model versions seamlessly.

\emph{Refactoring models} is the process of restructuring the model while not modifying its original functionality or behavior. This process can be automated using model transformation languages and refactoring libraries. Tools and frameworks, such as Epsilon~\cite{KolovosG22} or EMF Refactor~\cite{ArendtT12}, 
provide automated support for model refactoring.



\subsection{Automation-Enabling Techniques}
\label{enabling-technologies}

From the very beginnings of MDE, automation emerged as a fundamental pillar. Numerous works have employed direct approaches based on heuristics to automate or semi-automate certain activities directly on the models' representations. Other works have relied on established automation technologies to achieve this automation. In the remainder of this section, we discuss four families of these technologies: \textit{formal methods}, \textit{search-based techniques}, \textit{extensional and intensional knowledge engineering approaches}, and \textit{artificial intelligence}.

\subsubsection{Formal Methods}

Due to their rigor and precision, formal methods have been utilized to specify, analyze, and verify various MDE artifacts. The literature is abundant with contributions on the application of formal methods for automating various MDE tasks. However, we limit ourselves to a few examples of their most common uses. 

The first family of applications targets the analysis and verification of UML models. This involves techniques such as model checking~\cite{HilkenGBV18}, constraint satisfaction problem solving~\cite{EhrigKT09}, Boolean satisfiability~\cite{soeken2010verifying}, and rewriting logic~\cite{Duran22}. Another family of application of formal methods concerns the generation of models for metamodel or transformation testing. Here again, a variety of techniques have been explored, including typed graph transformation~\cite{AnastasakisBGR10}, equivalence partitioning~\cite{CabotCR07}, and satisfiability modulo theories~\cite{SemerathBLMV20}, to name a few. A third noteworthy family, although few works have covered it, is model completion. Its automation is achieved, for example, through first-order logic using Alloy~\cite{SenBV10}. Formal methods have also been applied to define the operational semantics of modeling languages, e.g., by using Maude and graph transformations~\cite{RiveraDV09,EngelsHHS00}. Formal methods have also been employed for model versioning~\cite{Westfechtel14,TaentzerELW14}, including diffing and merging, which have been managed in terms of graph modifications and graph transformations.

\ins{Although formal methods have contributed to explicitly modeling automation problems and provided effective solutions for some MDE task automation, their use is limited by several issues. The most well-known challenge is scalability. Formal methods often struggle when applied to large and complex models due to the computational cost of techniques like model checking and constraint solving. Moreover, certain MDE tasks, such as creative or exploratory modeling, are not easily automatable using solely formal methods. Finally, despite their potential, tools supporting formal methods often lack a direct integration with mainstream MDE environments, which hampers seamless workflows.}

\subsubsection{Search-based Techniques}

These techniques have emerged as valuable tools for automating various MDE activities. This line of research views the automation of many MDE tasks as optimization problems, which can be effectively tackled using search-based methods~\cite{boussaid2017survey}.
One of the earliest applications of this concept was in the domain of model transformations, particularly using examples of input-output models. Researchers utilized search-based techniques for model diffing based on refactoring detection\cite{KessentiniMWOD17}, model merging \cite{MansoorKLWBD15}, and to perform model transformations~\cite{kessentini2008model} and even to learn transformation rules~\cite{faunes2013genetic,baki2016multi}.

Moreover, search-based techniques have been instrumental in generating test suites to validate these transformations. Test cases, often in the form of input-output models, are generated to ensure the correctness of the transformations~\cite{jilani2014search}.
Additionally, model samples play a crucial role in testing metamodels. Search-based techniques have been employed to either generate those model samples~\cite{gomez2012searching} or to perform model selection~\cite{batot2016generic} to ensure the robustness and effectiveness of metamodels. In cases where metamodel descriptions lack well-formedness rules, these rules can be recovered from examples of both valid and invalid models. Genetic programming techniques have been utilized for this purpose, aiding in the refinement of metamodels~\cite{batot2022promoting}.

Another significant application of search-based techniques in MDE is repairing modeling artifacts. The evolution of metamodels can invalidate many associated artifacts due to their dependency on the type systems defined within the metamodels. Evolutionary algorithms have been employed to repair models~\cite{kessentini2022semi}, OCL constraints~\cite{batot2017heuristic}, and model transformations~\cite{kessentini2018automated}. Similarly and independently from the evolution, search-based approaches have been used to repair errors in MDE artifacts such as transformations~\cite{varaminybahnemiry2021automated}.

In the realm of design space exploration (DSE), where finding the best design instance for a given problem is challenging, search-based techniques have shown invaluable capabilities. DSE, framed as an optimization problem, involves exploring alternative models to identify those that meet specific criteria~\cite{abdeen2014multi}.

The relation between MDE and search-based techniques is also a tale of cross-fertilization. Indeed, MDE contributed to enriching search techniques through improved encoding of optimization problems and their subsequent resolution, e.g.,~\cite{DenilJVV14, BurduselZ018, JohnKLT23, BillFTMW19}.

\ins{In summary, search-based techniques have facilitated the automation of various MDE tasks by offering algorithms not only to solve specific instances but also to learn actionable knowledge that can be leveraged to streamline these tasks. However, this family of techniques has also several limitations. Tasks involving large search spaces are challenging to manage and often result in suboptimal solutions, particularly when determining the appropriate search parameters is difficult. Furthermore, the effectiveness of search-based techniques heavily relies on the quality of the objective functions guiding the search. Defining suitable, measurable, and well-balanced objectives within the context of MDE can be a significant challenge. Lastly, the solutions generated by search-based techniques may overfit specific problem instances, reducing their reusability and limiting their generalization to other MDE scenarios.}

\subsubsection{Extensional and Intensional Knowledge}

The automation or semi-automation of a task requires knowledge that can be operationalized through software tools. This knowledge may include elements that enable the direct execution of the task. It can also be exploited within a complex automation process. In the context of MDE, knowledge often appears extensionally in some kind of repositories, particularly in the form of sets of artifacts such as models~\cite{robles2017extensive} or transformations\footnote{\url{https://eclipse.dev/atl/atlTransformations}}. It can also be intentionally represented, typically through ontologies.

Among the artifacts within MDE, model sets have been extensively leveraged to provide support in modeling tasks. These collections of models are employed via search engines to propose models or model fragments aligning with specific criteria articulated by modelers, either in textual or pre-defined partial model forms~\cite{10.1145/2579991,lopez2022efficient, basciani2018exploring}.

The alternative category of knowledge, termed intensional, integrated into MDE task automation, encompasses ontologies. These structures provide organized descriptions of real-world objects, which can be utilized for crafting domain-specific languages or models~\cite{staab2010model}. Due to their inherent semantics, ontologies also serve in model verification and debugging processes~\cite{mokos2010ontology,walter2014ontology}, or for model matching~\cite{KappelKKKRRSW06}. Additionally, ontologies can be harnessed through design patterns to automate the creation of meta-models or models~\cite{AtkinsonK03a,pescador2016dsl}.

Lastly, the integration of extensional and intensional knowledge can enhance automation efforts. For instance, ontologies can streamline search requests and responses within model repositories~\cite{robles2012towards}.

\ins{Capturing and formalizing the knowledge required for automation, whether as artifacts or ontologies, is a complex and resource-intensive task. Automation solutions that rely on both types of knowledge are often highly domain-specific, limiting their generalizability and requiring significant effort to adapt the knowledge for use in new domains.
The effectiveness of extensional knowledge depends greatly on the quality, quantity, and relevance of the artifacts in the repository. Currently, existing repositories are often restricted to specific formalisms, such as UML class diagrams, making it challenging to assess the quality and representativeness of the artifacts. Additionally, maintaining large collections of models or transformations can become burdensome, with ensuring that repositories remain consistent and up-to-date posing a significant challenge.
Finally, although ontologies provide semantic richness, designing and maintaining them for complex domains is both time-intensive and requires specialized expertise in knowledge engineering.}

\subsubsection{Artificial Intelligence} 

Like many other disciplines, artificial intelligence has been extensively used as an automation technology for MDE. From the early days of MDE until now, rule-based systems have been employed to automate model transformations, utilizing either general-purpose rule languages such as NéOpus~\cite{revault1995metamodeling} or specifically dedicated ones like ATL~\cite{ATL}. Soon after, the notion of supporting modelers in writing transformation rules emerged. Machine learning techniques were applied to derive transformation rules from examples using supervised learning~\cite{varro2007automating} or unsupervised learning~\cite{dolques2009transformation}.

More recently, with advancements in machine learning research, considerable effort has been dedicated to assisting in metamodeling and modeling activities. This includes the utilization of techniques such as GNNs~\cite{di2021gnn} and LSTM~\cite{weyssow2022recommending}. Furthermore, research has been focused on model transformations using reinforcement learning~\cite{EisenbergPGW21} and LSTM architectures~\cite{BurguenoCLG22}. Various machine learning algorithms have also been applied for metamodel classification~\cite{nguyen2019automated} and model classification~\cite{lopez2022machine}, as well as for model repair~\cite{barriga2022parmorel} and model generation~\cite{lopez2022machine}. However, these initiatives often face challenges due to the difficulty of training machine learning models, particularly given the scarcity of data.

The remarkable power of Large Language Models (LLMs)\footnote{Note that a model in AI and a software model do not refer to the same type of model. In the AI field, a model is the result of analyzing datasets to find patterns and make predictions (e.g., it could be a trained neural network or a statistical model). To avoid confusion, in
this work, each time we refer to an AI model, we always refer to it with a precise name (e.g., AI model, ML model, Language model, ...) and never as ``model'' alone. When we use ``model'' alone, it always refers to a software model.} has partially alleviated this problem. Thus, automating several modeling tasks with little or no training data became possible. However, the effectiveness of using LLMs for this purpose varies, as Camara et al.~\cite{camara2023assessment} highlight. Ben Chaaben et al. explored the application of LLMs for static and dynamic domain-model completion~\cite{chaaben2023towards}. Model generation with LLMs has also garnered significant interest among researchers, encompassing diverse model types such as goal models~\cite{chen2023ontheuse} and static domain models~\cite{chen2023automated}.

\ins{Unlike other fields, the MDE research community has yet to fully capitalize on the potential of AI and, in particular, deep learning (DL) and large language models (LLMs) for automation, largely due to several inherent challenges. One major obstacle is the scarcity, fragmentation, and inconsistency of MDE-specific datasets, which are essential for effectively training and fine-tuning DL models. Moreover, MDE involves diverse, highly structured artifacts closely tied to domain-specific languages (DSLs). Adapting DL architectures to accommodate this diversity and to work seamlessly with domain-specific representations remains an unresolved challenge.
Another significant obstacle in leveraging LLMs for MDE tasks lies in the challenge of crafting effective prompts to extract relevant knowledge from the LLMs and mapping this knowledge accurately to the DSLs in which it must be represented. This complexity is compounded by the reliance of MDE practitioners on well-established toolchains. Integrating DL and LLM-based solutions into these ecosystems is challenging, particularly when existing tools are not designed to address the unique demands of MDE-specific tasks.}



\subsection{Mapping between Activities and Enabling Technologies}
\label{mapping}

The aim of this subsection is to present a mapping between the previously presented modeling activities and the enabling technologies. This mapping does not intend to be exhaustive by any means but wants to present illustrative examples. Table~\ref{tab:my-table}\footnote{Since a metamodel is a model (in particular, a model of a model), some of these activities apply to both models and metamodels. For the sake of brevity, we only refer to models.} shows activities in its rows and technologies in its columns (the column ``Direct'' contains examples of approaches providing automation support directly for the model representations without moving to an already existing automation space). Each cell contains a maximum of two exemplary works, appended with a `+' if the state of the art is more extensive and includes further works. Cells with only one work mean that we could not find more examples for that activity and technology. Empty cells mean that we could not find any example.

\begin{table*}[t!]
	\centering
    \begin{threeparttable}
   \resizebox{\linewidth}{!}{\begin{tabular}{|l|l|c|c|c|c|c|c|c|}
		\hline

		\multicolumn{2}{|c|}{\multirow{2}{*}{\textbf{\textit{Activities}}}} & \multicolumn{7}{c|}{\textbf{\textit{Enabling Technologies}}} \\ \cline{3-9} 
		\multicolumn{2}{|l|}{} & \textbf{Direct} & \textbf{Formal Methods}  & \textbf{Know. Eng.} & \textbf{Traditional AI\tnote{*}} & \textbf{Search Techniques} & \textbf{Deep Learning\tnote{**}} & \textbf{LLMs}\\ \hline

\multirow{4}{*}{\textbf{\textit{Modeling}}} & \begin{tabular}[c]{@{}l@{}} \textbf{Model generation}\end{tabular} & \cite{Gomez-AbajoGL16,BrottierFSBT06}+  & \cite{Varro2020,Lambers2017}+  &  & \cite{lopez2022machine} & \cite{jilani2014search,abdeen2014multi}+ &  & \cite{chen2023automated,chen2023ontheuse}\\ \cline{2-9} 
		& \begin{tabular}[c]{@{}l@{}}\textbf{Model search}\end{tabular} & \cite{Whittle2008,BascianiRRIP18}+ &  & \cite{lopez2022efficient}  &  &  &  & \\ \cline{2-9} 
		& \begin{tabular}[c]{@{}l@{}}\textbf{Model completion}\end{tabular} & \cite{Kuschke2013,BornBMPW08}+ & \cite{SenBV10,SteimannU13}+ & \cite{10.1145/2579991,BurguenoCGLC21}+ & \cite{10.1007/978-3-540-69489-2_18} &  & \cite{di2021gnn} & \cite{chaaben2023towards,tinnes2024leveraginglargelanguagemodels}\\ \cline{2-9} 
        & \begin{tabular}[c]{@{}l@{}}\textbf{Models at runtime}\end{tabular} & \cite{BencomoGFHB08,FouquetNMDBPJ12}+ & \cite{BurSVV18,SakizloglouGBG22}+ &  & \cite{BencomoBI13,0001MFT19}+ & \cite{KinneerGG21,EisenbergLSW22} &  & \\ \cline{2-9} 
	& \begin{tabular}[c]{@{}l@{}}\textbf{Model classification}\end{tabular} & \cite{RubeiRRNP21,CAISE2016}+ &  &  & \cite{nguyen2019automated,lopez2022machine} &  & \cite{NguyenRPRI21} & \\ \hline
  
        \multirow{2}{*}{\textbf{\textit{Model quality management}}} & \textbf{Analysis / Verification} & \cite{BascianiRRIP19,Ma_He_Liu_2013}+ & \cite{GogollaHD18,RiveraDV09}+ & \cite{mokos2010ontology} & \cite{barriga2022parmorel,WenzKT21}+ & & &\\ \cline{2-9} 
		& \textbf{Testing / Debugging} & \cite{PopoolaKR16,BagherzadehHD17}+ & \cite{HilkenGBV18,BabikianSLMV22}+ & \cite{walter2014ontology} &  & \cite{batot2016generic,gomez2012searching}+ & \cite{RahimiTRB23,LopezC23}+ & \cite{camara2024towards} \\ \hline
	  
        \multicolumn{2}{|l|}{\textbf{\textit{Model transformations}} }  & \cite{9146715,BiermannEKKTW06}+ & \cite{biermann2008precise,GammaitoniK19}+  & \cite{RoserB05,KappelKKKRRSW06}+  & \cite{EisenbergPGW21} & \cite{kessentini2008model,baki2016multi}+ & \cite{BurguenoCLG22,LanoX23} & \cite{Kazai_Osei_2024,Buchmann24} \\ \hline
		
  \multicolumn{2}{|l|}{\textbf{\textit{Model based execution}}} & \cite{MullerFJ05,CombemaleBW17}+ & \cite{RiveraDV09,EngelsHHS00}+ & &  &  &  &\\ \cline{2-9} \hline
  
		\multirow{4}{*}{\textbf{\textit{Model management}}} & \begin{tabular}[c]{@{}l@{}}\textbf{Model (co-)evolution}\end{tabular} & \cite{DIRUSCIO12,Herrmannsdoerfer_2011}+ & \cite{Taentzer_Mantz_Lamo_2012,LevendovszkyBNSBK14}+ &  &  & \cite{10.1145/3239372.3239375,kessentini2022semi}+ &  & \cite{JOT:issue_2024_03/article6} \\ \cline{2-9} 
		& \begin{tabular}[c]{@{}l@{}}\textbf{Model migration and maintenance}\end{tabular} & \cite{RoseKPPP14,HerrmannsdoerferR09}+ & \cite{GW06} &  & \cite{BarrigaMPRHI20} & & & \\ \cline{2-9} 
		& \begin{tabular}[c]{@{}l@{}}\textbf{Model versioning / diff / merge}\end{tabular} & \cite{KoegelHLHD10,EMFCompare08}+ & \cite{Westfechtel14,TaentzerELW14}+ & \cite{KappelKKKRRSW06} & & \cite{KessentiniMWOD17,MansoorKLWBD15}+ & & \\ \cline{2-9} 
		& \begin{tabular}[c]{@{}l@{}}\textbf{Model refactoring}\end{tabular} & \cite{SunyePTJ01,Porres03}+ & \cite{Mens06,BiermannEKKTW06}+ &  & & \cite{mokaddem2018recommending,GhannemKHE18}+ & & \\ \hline
	\end{tabular}}
    \begin{tablenotes}
  \item[*] \scriptsize{Excluding Search-based Techniques}
  \item[**] \scriptsize{Excluding LLMs}
  \end{tablenotes}
 \end{threeparttable}
	\caption{Overview of MDE activities and enabling technologies.}
	\label{tab:my-table}
\end{table*}

Table~\ref{tab:my-table} gives a summary of how automation-enabling technologies have been used for various MDE tasks.\footnote{Table~\ref{tab:my-table} is inevitably non-exhaustive and merely intended to reflect the kinds of activities and available supporting approaches.}  Many techniques, including direct programming, formal methods, and search methodologies have been extensively employed to automate various MDE tasks. To a lesser extent, the MDE community has leveraged, on the one hand, model sets, and ontologies and, on the other hand, other AI techniques to automate some distinct activities. On another note, in recent years, there has been a surge in interest and adoption of deep learning-based approaches for MDE. However, the scarcity of large-scale datasets for model training or refinement has hindered their widespread use.

The utilization of Language Model-based systems and other cutting-edge technologies in this realm is still in its nascent stages. To date, the primary focus has been on model generation and completion tasks within MDE workflows. Nonetheless, this landscape presents numerous untapped research directions for further exploration and refinement of deep learning approaches in MDE.

To capitalize on these opportunities, there is a pressing need to prioritize the creation of expansive datasets tailored specifically for MDE tasks, as highlighted in Section \ref{sec:perspectives}. By doing so, researchers and practitioners can unlock the full potential of deep learning methodologies in automating and enhancing various tasks of MDE.

Looking ahead, LLMs are poised to become a pivotal research domain in automating MDE tasks. However, their application extends beyond mere model generation or completion tasks. A vast scope exists for creative exploration and utilization of LLMs in diverse MDE activities, heralding a new era of innovation and efficiency in software engineering practices.

\section{Perspectives and Challenges}
\label{sec:perspectives}

Building upon the introductory concepts discussed earlier, this section provides an overview of perspective topics that hold significance in the medium to long term for automation in MDE. We foresee a vast potential for automation within MDE, driven by advancements in AI, particularly in generative AI and LLMs. In exploring the future prospects of MDE and its automation possibilities, it is crucial to acknowledge the challenges ahead. Challenges include integrating automation tools, ensuring compatibility, and maintaining interoperability among diverse modeling languages and environments. Additionally, developing intuitive modeling tools and addressing ethical considerations in interdisciplinary domains pose significant problems to be investigated, as discussed below. The distinction between perspectives and challenges may not always be clear-cut and may overlap. While discussing envisioned advancements, we see some of them as potential obstacles for automation that must be overcome.

\subsection{AI and MDE}
\label{AI-MDE}

As with many disciplines, AI significantly enhances the potential for automating tasks. In the past decade, considerable advances have been made, particularly in natural language and code models. MDE is ideally positioned to benefit from both types of LLMs, as it uniquely represents software at a high level of abstraction while also translating these abstractions into operational implementations. However, these opportunities come with many challenges, as AI models have not been specifically designed for these uses, unlike code models, whose design is aligned with coding tasks. In this section, we discuss these opportunities and challenges.

\smallskip
\noindent
\textit{Combination of symbolic and non-symbolic AI}: Integrating symbolic AI techniques—such as logic reasoning and rule-based systems—with non-symbolic approaches like neural networks and deep learning presents significant opportunities for advancing automation in MDE. By harnessing the complementary strengths of these paradigms, researchers can develop hybrid AI systems that excel in tasks requiring both symbolic reasoning and pattern recognition. For instance, combining symbolic reasoning with neural networks allows MDE tools to interpret and manipulate models according to logical constraints, while leveraging the scalability and adaptability of neural networks for data-driven tasks. Additionally, generative AI tools can synthesize domain-specific or context-aware models, further enhancing the effectiveness and applicability of MDE tools.

\smallskip
\noindent
\textit{Emancipation from formal languages with LLMs}: Creating domain-specific modeling languages is a critical endeavor that requires collaboration among various roles, including language and domain experts. This process involves several key steps: developing concrete syntaxes, defining formal semantics, and constructing supportive modeling environments. However, the language development process often presents challenges, such as refining language definitions to meet evolving user requirements or address unforeseen issues. Additionally, the resulting languages may not always be user-friendly for domain experts. LLMs offer a promising solution for the early evaluation of these languages. By specifying their semantics in natural language, LLMs can facilitate preliminary analysis and stakeholder interactions, helping to ensure that the language aligns with user needs. This approach can be further extended by using natural language to represent different aspects of a domain and allowing LLMs to manage ambiguity in these representations for subsequent development tasks.

\smallskip
\noindent
\textit{Moving from general LLMs to task-specific AI agents:} In the traditional MDE context, a platform typically refers to a specific technology stack or execution environment. However, with the growing adoption of LLMs, the concept of a platform broadens to encompass a collaborative ecosystem of multiple language models operating within a multi-agent system. We anticipate a paradigm shift toward the widespread use of task-specific AI agents powered by LLMs. These specialized agents are designed to handle various model management tasks, including model evolution, comparison, domain modeling, training and test data generation, and model completion \cite{hong2023metagpt}.

\smallskip
\noindent
\textit{Orchestration of automated MDE tasks}: Building on the previous point, applying modeling concepts and tools involves performing various tasks, such as model transformations, validation, and code generation. Despite the availability of domain-specific languages like Wires \cite{RiveraRLB09}, there remains a need for higher-level automation strategies that can, starting from user requirements, orchestrate low-level automated tasks while remaining flexible enough to support different modeling languages, tools, and environments. Furthermore, achieving seamless integration of these automated processes requires overcoming compatibility issues and ensuring interoperability between diverse automation tools and frameworks.

\smallskip
\noindent
\textit{Model management automation with AI approaches}: AI-driven techniques can significantly enhance model management processes by improving the semantic understanding and context-aware manipulation of models. By applying AI to tasks such as model differencing, merging, and versioning, MDE can achieve higher efficiency and reliability in handling evolving model artifacts. For instance, AI algorithms can analyze the semantics of models to detect and resolve—or propose solutions for—conflicts during model merging, reducing manual effort and minimizing errors in model integration. 

\smallskip
\noindent
\textit{AI to improve reuse in MDE}: Unlike in the realm of code, the level of artifact reuse in MDE remains low. This can be partly attributed to the limited presence of explicit semantics associated with these artifacts and the vast and ambiguous vocabulary often employed. However, the ability of LLMs to grasp the meaning of representations unlocks new possibilities for reuse. This includes enhanced model search, model completion, and even the generation of model transformations.

\smallskip
\noindent
\ins{
\textit{Template-based code generation combined with AI-based code generation}: While template-based methods provide transparency and control over the generated code, AI-based approaches offer adaptability and potential for optimizing code generation based on diverse requirements and contexts \cite{weyssow2024codeultrafeedback}. Researchers can develop code generation techniques by exploring hybrid approaches that combine the strengths of both paradigms. Also, for emerging platforms, an important research challenge is how code generation facilities can be provided as reference applications from which code generation schemes may be abstracted. This challenges both paradigms, which may be mitigated by having hybrid code generators that can resort to human-defined templates combined with AI-based code injections to balance the required human efforts as well as the demand for large sets of reference applications. 
}

\smallskip
\noindent
\textit{Scarcity of datasets and benchmarks:} 
The scarcity of high-quality datasets hinders the advancement of AI-driven automation in MDE and benchmarks for training and evaluating AI models. Unlike domains such as natural language processing or computer vision, where large-scale datasets are readily available, MDE lacks comprehensive datasets that capture a diverse range of modeling tasks and application domains \cite{Izquierdo_Cuadrado_2024}. In addition, concerns about data privacy and the ethical use of data further complicate the acquisition and sharing of modeling datasets. Addressing these challenges requires collaboration between researchers, practitioners, and stakeholders to curate datasets, define evaluation metrics, and establish best practices for data collection and sharing in MDE contexts \cite{RoccoRIP15,camara2024towards}. 
 
The advent of LLMs presents new opportunities for the creation of datasets in various domains. With the ability to generate human-like text and simulate expert knowledge, LLMs offer a cost-effective alternative to traditional human labeling methods. For instance, LLMs can automatically generate labeled datasets for MDE tasks such as model classification or sentiment analysis. Additionally, LLMs can serve as judges to evaluate the quality and accuracy of tasks performed by other LLMs, thereby facilitating continuous improvement and refinement of AI-driven systems.

\subsection{Human Factors and MDE\del{ (including education)}}
\label{human-factors}

Models enable communication among stakeholders, software engineers, and final users. Models and MDE also increase the productivity of software development teams thanks to the (partial) automation of the development process~\cite{2012Brambilla}.
When using models, humans have goals and expectations, which should be met as both the success of their projects as well as the modeling disciple depend on this. Therefore, special attention should be paid to how humans understand and interact with models. As MDE processes are being automated and the field is evolving, exploration of how these changes affect humans needs to be explored. It is also worth noting that Liebel et al. have reported that ``there is an underrepresentation of research focusing on human factors in modeling''~\cite{LiebelKHLNGSEOMJSGHHWTG24}.
In the following, we discuss the opportunities and challenges that the latest advances in MDE are bringing to the field.

\smallskip
\noindent
\textit{Teaching MDE:} Given the rapid evolution of technology and MDE, teaching MDE needs to be adapted to the fast evolving needs. Educators must prepare students not just to be proficient modelers, but to be architects of highly complex software and systems that leverage automation and MDE techniques to meet the demands of industry and society\ins{~\cite{BucchiaroneCPP20}}. Incorporating new technologies into the teaching of MDE can significantly improve the learning experience. For instance, the use of simulation tools, multidimensional visualization software, and other interactive applications can provide a more immersive and engaging learning experience.

\smallskip
\noindent
\textit{Teaching people how to prompt \ins{for MDE automation:}} Despite advances in AI and automation, human experience remains indispensable to designing complex systems and formulating effective prompts for AI models. \ins{In the case of MDE, these prompts or pipelines of prompts) need to take into consideration different aspects of MDE artifacts such as syntax, semantics, abstraction level, etc. (e.g.,~\cite{SipioRRRI24,dirocco2024uselargelanguagemodels}} deal with these problems). Teaching people how to prompt effectively \ins{for MDE} involves imparting knowledge and skills in understanding system requirements, defining clear objectives, and structuring \ins{and designing} prompts to elicit desired responses from AI models. Thus, human expertise is essential for ensuring the accuracy, relevance, and ethical considerations of AI-generated outputs. 

\smallskip
\noindent
\textit{Combination of physical/biological/social/software aspects in complex systems:} Integrating physical, biological, social, and software aspects into the same system presents a multifaceted challenge that requires a comprehensive approach to modeling and abstraction. MDE stands as an ideal integration vehicle for managing these heterogeneous aspects by providing a unified framework for representing and analyzing complex systems. Digital twins, which are virtual replicas of physical entities or systems, offer a powerful abstraction mechanism for simulating and analyzing the behavior of interconnected systems. By leveraging digital twins, MDE can facilitate the design and optimization of integrated systems across diverse domains\ins{~\cite{bordelau20}}, ranging from healthcare to sustainable smart cities.

\smallskip
\noindent
\textit{Emergence of non-traditional interface devices}: Integrating speech interfaces, gesture recognition, and other non-traditional input modalities in MDE tools can democratize software development by allowing a broader range of stakeholders to participate in modeling activities\ins{~\cite{EloualiRPT13,PlanasDBC21}}. By providing intuitive and accessible interfaces, such as voice commands or touch-based interactions, MDE tools can lower entry barriers for users with diverse backgrounds and abilities. Moreover, non-traditional interface devices facilitate collaboration and communication among distributed teams, allowing users to interact with models in real time and co-create solutions collaboratively.  To this end, interdisciplinary work at the intersection of the fields of human-machine interaction and software engineering research~\cite{Kashfi_Feldt_Nilsson_2019}. 

\smallskip
\noindent
\textit{Democratizing the usage of task-specific AI agents:} To enable the usage of task-specific AI agents, it is necessary to devise low-code platforms that \ins{prioritize human-centric design, that is, that} allow users to take advantage of task-specific AI agents seamlessly.
\ins{Such platforms can facilitate broader participation in MDE activities across domains and expertise levels by providing intuitive interfaces for coordinating diverse AI models, empowering users from different backgrounds and expertise to interact effortlessly with AI agents}. In addition, low-code platforms can serve as orchestrators for deploying and managing AI agents, allowing users to customize and extend automation capabilities according to their specific requirements and preferences.

\smallskip
\noindent
\textit{Feedback-driven MDE:} MDE tools and processes will need built-in feedback mechanisms. This feedback could come from engineers, stakeholders, or the system itself (in the form of real-time data) and could be sporadic or continuous. Tools, processes, and teams should not disregard feedback. New methodologies are needed to work in new environments where automated feedback integration is supported~\cite{KienzleCMABBEGJ22}.
A straightforward scenario is Digital Twins. For instance, Digital Twins can be used for product development, where design and engineering teams can investigate a broader range of design possibilities early in development, saving money, resources, and time. In such a scenario, the engineering processes must be adapted to collect and react to the outputs/feedback received by the Digital Twin, and engineering teams must have methodologies put in place.

\smallskip
\noindent
\textit{Automated and semi-automated decision-making in MDE:} As software and systems grow in complexity and the demands for faster delivery increase, automating decision-making processes becomes necessary to improve efficiency and quality. On a purely technical level, automated decision-making will help choose the best transformation languages, model checkers, AI algorithms, execution models, etc. Furthermore, it would ensure that decisions are made consistently throughout the life of a project and by different teams. It would also prevent human error in repetitive decision-making processes (e.g., tools enhanced with automated support could exploit data and human-defined objectives to make or suggest optimal decisions).


\subsection{\ins{Diversity of Application Domains, Users, and Modeling Interfaces}}
\label{modeling-interfaces}

MDE has been extensively explored for classical software systems such as embedded systems, cyber-physical systems, Web-based systems, distributed systems, etc. The literature also discusses that MDE shows different adoption levels~\cite{vallecillo2015industrial} in such domains due to organizational or technical opportunities/challenges. However, in recent years, we also face several new computing paradigms emerging such as quantum computing, neuromorphic computing, extreme parallel computing, intensive data-driven computing, etc. In the future, it has to be explored if MDE-based automation is also possible and beneficial for these new computing paradigms or if new MDE modeling interfaces are required for these emerging computing paradigms before MDE can be applied to them effectively. In addition, new interaction possibilities based on emerging technologies are possible with models, which opens innovative ways how models are constructed, maintained, and used in diverse contexts and users, going beyond what is known nowadays from typical standalone desktop-based modeling editors. In the following, we discuss several opportunities and challenges in this respect.

\smallskip
\noindent
\textit{Management of ever-increasing diversity of languages and tools:} MDE has seen a proliferation of domain-specific modeling languages and tools, which has led to a diverse range of technologies. However, this diversity has posed challenges for \emph{generic} automation as each modeling language has its own syntax, semantics, and transformations. Despite successes in standardizing the syntax of models, e.g., see the standardized metamodeling language MOF\footnote{\url{https://www.omg.org/mof}} which is frequently used for MDE technologies, the MDE field still needs more industry-relevant standards, especially concerning semantics, transformations, and model interchange formats. Moreover, the rise of low-code/no-code platforms makes this issue even more difficult as adherence to existing standards becomes increasingly sporadic~\cite{RuscioKLPTW22}. Thus, it is imperative to develop standardized interfaces, transformation mechanisms, and model interchange formats to ensure interoperability and tooling support across different formalisms, languages, and technologies. This perspective is even more challenged by also supporting new computing paradigms which may require completely new standards or the extensive adaptation of existing ones. 

\smallskip
\noindent
\textit{Multi-modal modeling tools:} Despite significant advancements in modeling tools, many existing tools in MDE are tailored towards expert users (who know the intrinsics of the modeling languages and associated tools very well) but lack intuitive interfaces for non-specialists. This poses challenges for automation and democratizing the adoption of MDE methodologies across diverse domains and user communities~\cite{RuscioKLPTW22}. Improving the accessibility~\cite{KHALAJZADEH2025107570} and usability~\cite{ChakrabortyL24} of modeling tools requires incorporating user-centered design principles, conducting usability studies, and providing comprehensive documentation and training materials. Additionally, ensuring compatibility and interoperability between different modeling tools and environments is essential for facilitating seamless collaboration and exchange of models among users with varying expertise levels. This is also of paramount importance for supporting emerging computing paradigms~\cite{AliY20} and novel human-computer interaction paradigms~\cite{SunkleSPK22}, as it would allow to involve experts from different disciplines which have to be supported in the best possible way to let them participate in the development process. Again, the trade-off between having a smooth modeling process and having high quality models has to be explored in this setting to ensure a good level of automation.

\subsection{Modeling and non-functional concerns}
\label{non-functional-concerns}

\ins{Non-functional requirements play a pivotal role in software engineering, impacting aspects of software quality beyond functional correctness. In the context of MDE, the research community has been focusing on different quality aspects, such as the \textit{performance} of model transformations \cite{10.1145/3578244.3583727}, \textit{scalability} of model management tools \cite{BruneliereKDMKC20}, and \textit{maintainability} of modeling artifacts to enhance their reuse possibilities \cite{DBLP:journals/spe/IndamutsaRARP24}. Managing the \textit{(co-)evolution} of modeling artifacts has also been the subject of intense research \cite{DIRUSCIO12}, and only recently, the introduction of alternative modeling means, e.g., virtual reality \cite{10.1016/j.infsof.2023.107369} or augmented reality \cite{JOT-issue_2023_02/article7}, has also been an interesting research topic.}
Critical concerns, such as \textit{security}, are well-recognized in modern software systems due to the increasing threats posed by cyber-attacks and vulnerabilities in complex systems. At the same time, newer emerging concerns, such as \textit{sustainability}, are rapidly gaining importance as software systems evolve, and it is necessary to ensure they meet the demands of energy efficiency, resource optimization, and environmental impact reduction. Both of these concerns require research efforts in the MDE community, which has provided significant results in the automation of various software engineering and development tasks. Thus, it is necessary to devise techniques and tools to make MDE automation approaches aware of security and sustainability concerns. 

In the following, as examples of \textit{existing} critical concerns and \textit{emerging} ones, we emphasize research challenges and future directions related to managing security and sustainability in the context of MDE automation.

\smallskip
\noindent
\textit{Security management:} A critical future research direction in MDE automation is the management of security aspects of complex software systems in critical fields like IoT and cyber-physical systems. In this respect, MDE plays a crucial role because it can enable dynamic security adaptations, threat modeling, and the generation of test cases for security policies. Recently, Riegler et al. \cite{Riegler2023} proposed a model-driven approach to manage multi-modal system architectures, allowing for automated mitigation strategies through configurable mode switches. A dedicated language for modeling mitigation strategies enables such automation. However, advances in model-based security verification, co-evolution, and access control policies will be essential for enhancing both system security and the security of MDE tools themself. 

As generative AI technologies become more prevalent in supporting software engineering tasks, it is increasingly important to develop techniques and tools that ensure code security, which is obtained through model-based generators. Thus, research efforts are needed to integrate security checks and validation techniques within MDE automation tools. One promising direction is the exploration of model alignment techniques, which can help refine outputs generated by LLMs in line with user expectations. For instance, Ganqu C. et al.
\cite{cui2024ultrafeedbackboostinglanguagemodels} proposed an approach to align LLM outputs with human preferences, learned from human feedback. Applying similar model-alignment strategies could enhance the code generation capabilities of LLMs by detecting and flagging malicious or security-sensitive code. This feedback loop would prevent the generation of such unsafe code in future iterations.


\smallskip
\noindent
\textit{Sustainability management:} Empirical evidence from the industry highlights the challenges posed by limited support for managing large models, which hampers the broader adoption of MDE in industrial applications \cite{hutchinson2014model}. The size of these models can also negatively affect runtime performance, causing delays during MDE operations such as model transformations \cite{robles2014improving}. Significant research has been conducted on improving scalability to address these issues, with solutions like the Gremlin framework specifically designed to mitigate these limitations \cite{daniel2017gremlin}. Moreover, recent work has highlighted the persistent performance challenges in model transformations \cite{burgueno2019future} and suggested potential enhancements through static analysis \cite{9146715}. Despite these advances, energy consumption has been largely neglected in discussions around MDE scalability. With data centers projected to consume 10\% of global electricity by 2030 \cite{verdecchia2021green}, there is a growing recognition of the importance of energy efficiency by academia and industry \cite{fonseca2019manifesto}.

\section{Conclusion}

MDE has proven its utility over time, primarily due to its remarkable capabilities in abstraction and automation. These features have significantly streamlined the development process, from conceptualizing ideas to their implementation. This article delves into the various facets of MDE activities and the diverse technologies that have been deployed to automate these activities.

The evolution of MDE is closely intertwined with the emergence of cutting-edge technologies, such as deep learning and large language models. These advancements have revolutionized automation in MDE activities, paving the way for enhanced efficiency and effectiveness in software development. In addition to offering extraordinary automation capabilities for MDE, these emerging technologies create new opportunities for this paradigm to play new roles in development processes. 
One significant aspect to consider is the democratization of AI techniques. The power of abstraction and automation provided by MDE allows non-specialists to leverage AI capabilities seamlessly. This democratization can lead to more widespread adoption of advanced technologies and contribute to innovation across industries.
Moreover, MDE's potential extends beyond traditional software development. It can be leveraged to build complex and hybrid systems that integrate physical, biological, social, and software components. Such systems are becoming increasingly crucial in modern societies, where interconnectedness and integration are paramount. \ins{It is important to remark that a critical gap in current MDE research is the limited availability of artifacts, such as datasets, benchmarks, and systematic evaluations, to validate tools, technologies, and methodologies, see e.g.,~\cite{BollVK24} for a recent study. This gap becomes even more significant with the increasing adoption of AI for MDE automation, where empirical evidence is essential to demonstrate the effectiveness, reliability, and scalability of AI-driven approaches. Future research should prioritize the creation of open datasets, developing standardized benchmarks, and establishing rigorous evaluation frameworks to ensure the practical applicability of MDE automation in different settings.}

\section*{Acknowledgment}
This work was partially funded by the Spanish Government (Ministerio de Ciencia e Innovación--Agencia Estatal de Investigación) under projects PID2021-125527NB-I00, TED2021-130523B-I00; by the Universidad de Málaga under project JA.B1-17 PPRO-B1-2023-037; and by the NSERC grant RGPIN-2019-07168. Additionally, this work was partially supported by the following Italian research projects: EMELIOT (PRIN 2020, grant n. 2020W3A5FY) and TRex SE (PRIN 2022, grant n. 2022LKJWHC). Additional support was provided by the European Union NextGenerationEU through the Italian Ministry of University and Research for the MATTERS project, funded under the cascade scheme of the SERICS program (CUP J33C22002810001), Spoke 8, within the Italian PNRR Mission 4, Component 2, as well as the FRINGE project (PRIN 2022 PNRR, grant n. P2022553SL).

\bibliographystyle{ACM-Reference-Format}
\bibliography{library}


\end{document}